\documentstyle[twocolumn,epsf,aps]{revtex}
\draft
\begin{document}
\title{Phase transition in a spatial Lotka-Volterra model}
\author{Gy\"orgy Szab\'o$^1$ and Tam\'as Cz\'ar\'an$^{2,3}$}

\address
{$^1$Research Institute for Technical Physics and Materials Science \\
POB 49, H-1525 Budapest, Hungary \\
$^2$Theoretical Biology and Ecology Research Group of the Hungarian
Academy of Sciences \\
$^3$Department of Plant Taxonomy and Ecology, E\"otv\"os University \\
H-1083 Budapest, Ludovika t\'er 2, Hungary}

\address{\em \today}

\address{
\centering{
\medskip \em
\begin{minipage}{15.4cm}
{}\qquad Spatial evolution is investigated in a simulated system of nine competing
and mutating bacterium strains, which mimics the biochemical war among bacteria
capable of producing two different bacteriocins (toxins) at most.
Random sequential dynamics on a square lattice is governed by very
symmetrical transition rules for neighborhood invasions of sensitive strains
by killers, killers by resistants, and resistants by sensitives. 
The community of the nine possible toxicity/resistance types undergoes
a critical phase transition as the uniform transmutation rates between
the types decreases below a critical value $P_c$ above which all the nine
types of strain coexist with equal frequencies. Passing the critical mutation rate
from above, the system collapses into one of three topologically
identical (degenerated) states, each consisting of three strain types.
Of the three possible final states each accrues with equal probability
and all three maintain themselves in a self-organizing
polydomain structure via cyclic invasions. Our Monte Carlo
simulations support that this symmetry-breaking transition
belongs to the universality class of the three-state Potts model.
\end{minipage}
}}
\maketitle

\narrowtext

Many species of bacteria have recently been shown to excrete toxic
subtances that are very effective against strains of the same or closely
related species not producing the corresponding resistance factor
\cite{bacik1,bacik2}. With respect to a certain toxin a species consists of
colonies of three possible types: Sensitive ($S$), Killer ($K$), or
Resistant ($R$). Killer strains produce the toxin and a resistance factor
that prevents suicide; resistant strains only produce the resistance factor,
and sensitives produce neither. An $S$ colony can always be invaded and
displaced by a $K$ propagulum because $K$ can kill $S$ using the toxin.
A $K$ colony can be invaded and ultimately displaced by an $R$ propagulum
because the resistant type is immune to the toxic effect and it does not
carry the metabolic burden of synthetizing the toxin, thus it achieves a higher
growth rate and competitive dominance over $K$. The sensitive ($S$)
type can in turn displace the resistant by competition, for it does not
even pay the metabolic cost of producing the resistance factor. The
resulting cyclic pattern of competitive dominance ($K$ beats $S$ beats $R$
beats $K$) is a striking realization of the well-known Rock-Scissors-Paper
game \cite{HS}, by a biological entity. Other cyclic dominance systems
are almost unknown in ecology, but given the extraordinary significance of 
bacterial communities in virtually all ecological systems, the problem is 
well worth detailed theoretical studies. 

Some theoretical aspects of cyclic dominance have already been thoroughly 
investigated \cite{BG,HS}. In the simplest spatial (lattice) version
of a cyclic Lotka-Volterra system \cite{Lotka,Volterra}
the species are distributed on a $d$-dimensional lattice, and invasions
are confined to nearest neighbor sites with uniform rates.\cite{T94,FKB}
Analytical and numerical calculations have proven that fixation occurs
if the number of species exceeds a critical value dependent on dimension
$d$, otherwise a self-organizing domain structure is maintained for
$d \geq 2$.\cite{FK}, which comprises rotating vortices and antivortices 
in three-species models.\cite{T94,SSM}
The present work is meant to demonstrate that extending the cyclic 
dominance approach to a two-toxin bacterial community with mutation results 
in a remarkable enrichment of interesting dynamical phenomena, compared to what 
is already known.

The relevant biological details of bacteriocin systems are the following: 
The genes coding for the toxin and the resistance factor are usually both
sitting on an extrachromosomal DNA-ring in the citoplasm (called plasmid)
that the bacterium can lose and obtain without any immediate deleterious impact.
Each of the two genes on the plasmid can be switched off by DNA
mutation. Thus all possible mutational transformations are possible
in principle, but---supposing the mutant does not disperse far
immediately---those having a visible effect are only the ones after which
the mutant defeats the resident strain from which it emerged.
Obviously, $S \to K$, $K \to R$, and $R \to S$ are mutations followed
by competitive displacement of the resident, i.e., they are permitted,
but the reverse ones are immediately eliminated by the resident population.
$S\to K$ involves obtaining a complete plasmid which is possible, e.g., through 
genetic transformation or a sexual event called conjugation; the other two 
viable mutations are realized by switching off the toxin gene and the resistance
gene, respectively, on an existing plasmid.

\begin{figure}
\centerline{\epsfxsize=7cm
            \epsfbox{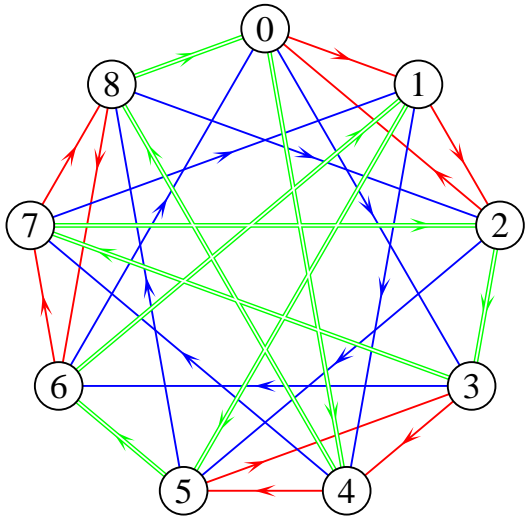}
            \vspace*{2mm}   }
\caption{Topology of dominance in the nine-species model.
The single and double lines with arrow indicate single and double
dominances as described in the text.}
\label{fig:foodweb}
\end{figure}

Most bacteria are capable of producing more than one toxin and/or the
corresponding resistance factors simultaneously. If the maximum number
of toxin types is two, the number of possible toxicity/resistance
combinations in a strain is nine. These are: $SS$, $SK$, $SR$, $KS$,
$KK$, $KR$, $RS$, $RK$, and $RR$. Here we confine our attention
to this two-toxin system, denoting the actual state of a bacterium
colony by an index number from $0$ to 8 in the order above. The topology
of the dominance relations among the states is illustrated in
Fig.~\ref{fig:foodweb}. The biological justification for this topology 
is straightforward: double dominance of strain A above strain B means 
that A harbors dominant genes on both plasmids compared to B; single 
dominance means that one gene of A is dominant, the other is identical 
to that of the correspondig gene in B; no dominance follows either if 
the corresponding genes are both identical, or if the two genes play 
draw (i.e., A wins with one gene, and B with the other).

We use a square lattice of size $L \times L$ with periodic boundary conditions 
as the arena for interaction. Each lattice site $i$ is occupied
by a colony of one of the nine toxicity/resistance types (we call them 
"species" for brevity in the sequel). Assignment of a certain value of the
state variable $s_i=0, \ldots , 8$ to a certain site indicates the presence
of the corresponding species on that site.
The simulation works by iterating the following elementary steps: At a 
randomly chosen site one of the two possible mutants replaces the resident
colony with probability $P$, otherwise the resident colony fights with a
randomly chosen nearest neighbor colony by mutually invading propagules
(with a probability $1-2P$). The outcome of the battle between two neighbors
($i$ and $j$) depends on the dominance relations between them: $i$ displaces
$j$ and takes over its site if $s_i$ is dominant over $s_j$ (cf.\ 
Fig.~\ref{fig:foodweb}). If the neighbors are equivalent or neutral to each
other (play draw), nothing happens.

Lattice size varies from $L=400$ to 3000 in the simulations. At $t=0$ the 
species were distributed at random with uniform probabilities on the lattice
in all simulation runs. The control parameter was mutation rate, varying
from 0 to 1/2. We have recorded the time series of species concentrations and 
correlation functions, the latter of which served as the basis for calculating 
correlation length. After a suitable thermalization time we have averaged these
data over some sampling time chosen to be long enough for providing sufficient
accuracy in spite of the occassionally very high concentration fluctuations.

In no-mutation runs ($P=0$) we have observed an interesting domain size increase
phenomenon in the time series of spatial patterns which develop. One can
distinguish three equivalent types of growing domains consisting of the 0+4+8,
1+5+6, and 2+3+7 species respectively. Inside these three domains a
self-organizing structure is maintained through the mechanism described by
Tainaka for the simplest 3-species cyclic dominance model \cite{T94}. We call
these domains "alliances" henceforth, but note that the species 
within an alliance are in fact the worst enemies: each alliance consists of
species cyclically double-dominating each other. Our reason for this terminology
will be clear in a minute.

The three alliances are given approximately equal territories at start, but the 
system slowly drifts towards a single-domain state in all simulations. Each
alliance has the same chance to take over.

Alliances defend themselves against the external invasion of "alien" species
with a peculiar mechanism. One can easily check that any external invador can
attack only one of the species within an alliance, and the invador is eliminated
from the domain of the alliance by the species actually controlling the attacked
one within the alliance. This means that the invador is wiped away by the toughest
within-domain enemy of the attacked species, thus maintaining the self-organizing
structure and the integrity of the domain with the very same mechanism.

\begin{figure}
\centerline{\epsfxsize=8cm
            \epsfbox{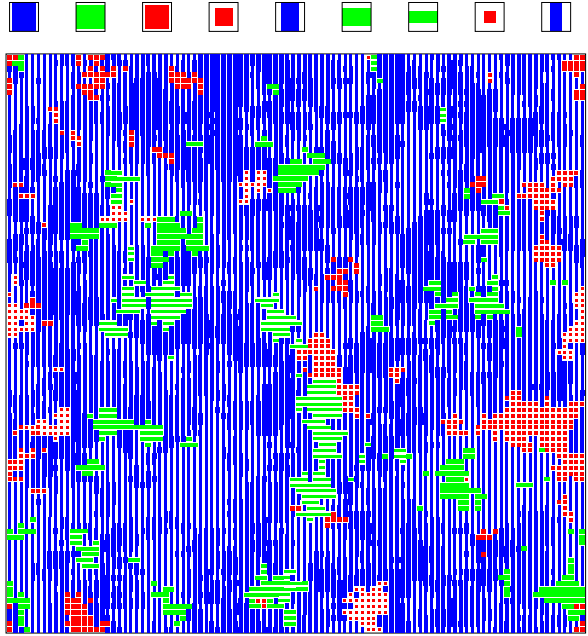}
            \vspace*{2mm}   }
\caption{Snapshot of species pattern for $P=0.0003$. The different
symbols magnified at the top represent the species $s=0, \ldots , 8$ 
from left to right.}
\label{fig:cfgordrd}
\end{figure}

The self-protection of alliances against external invadors can also be observed
for small mutation rates as illustrated in Fig.~\ref{fig:cfgordrd}. In this
snapshot similar symbols are used for species belonging to the same alliance.
Namely, different strip widths (or box sizes) distinct the species within the
three alliances represented by horizontal and vertical strips and closed
squares respectively. This figure illustrates that the mutants and their
offspring can form only small, temporary islands in the sea of the dominant
domain (0+4+8) represented by vertical strips. Clearly, the concentrations
of minority species increase with mutation rate $P$.

The average concentrations of the species become equal if the mutation
rate exceeds a critical value $P_c$. This continuous transition is
accompanied by a divergence in both the fluctuations and the correlation
length. A similar phase transition occurs in the three-state Potts
model \cite{Potts,Wu}, therefore we have adopted the numerical techniques
suggested by Binder \cite{Binder} for the quantitative analyses.

In order to reduce relaxation time in systematic investigations, the random
initial state contained only species 0, 4, and 8 at small values of $P$.
The MC simulations were performed on large lattices ($L>1500$) and long
sampling times ($t>3 \cdot 10^5$ MC steps per sites) in the vicinity of
the critical point. For such large lattice sizes the dominance of the
0+4+8 alliance was maintained during the simulations at all tested values
of $P$ below $P_c$.

\begin{figure}
\centerline{\epsfxsize=8cm
            \epsfbox{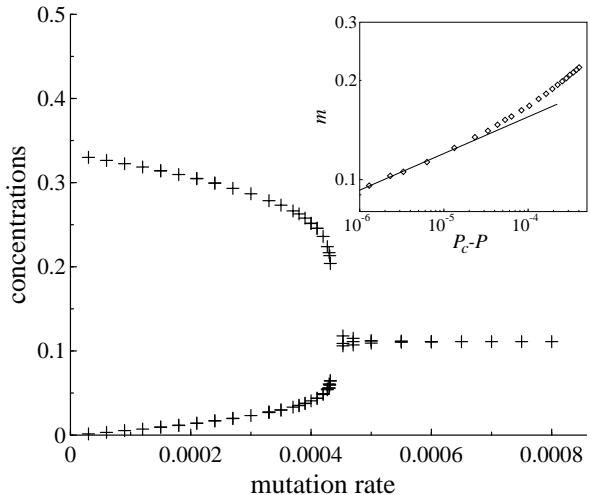}
            \vspace*{2mm}   }
\caption{Average concentrations of species as a function of mutation rate.
The inset shows the log-log plot of the order parameter vs. $P_c-P$.}
\label{fig:c}
\end{figure}

The simulations suggest a continuous transition as indicated in
Fig.~\ref{fig:c}. Statistical errors are small, comparable to the line
thickness of the figure, except for the very close vicinity of the
critical point. 

Figure \ref{fig:c} shows that the $P$-dependence of average concentrations
can be characterized by a single order parameter $m$ as
\begin{eqnarray}
c_0&=&c_4=c_8={1 \over 9}+m \, , \\
c_1&=&c_2=c_3=c_5=c_6=c_7={1 \over 9} -{1 \over 2}m \, . \nonumber
\label{eq:op}
\end{eqnarray}
According to our MC simulations, $m$ follows a power law behavior in the
close vicinity of $P_c$, i.e.
\begin{equation}
m \propto (P_c-P)^{\beta}
\end{equation}
if $P<P_c$ (see the inset in Fig.~\ref{fig:c}), whereas the order parameter
remains zero for $P>P_c$.
Numerical fitting yields $\beta = 0.110(5)$ and $P_c=0.0004333(5)$. This
value of the $\beta$ exponent is in good agreement with the theoretical
prediction ($\beta = 1/9$) obtained for the three-state ($Q=3$) Potts
model \cite{Wu}. This is not surprising, given that a large class of
two-state dynamical systems exhibits phase transition belonging to the
universality class of the Ising model \cite{Grins} and the $Q$-state Potts
model was introduced as a generalization of the Ising model \cite{Potts,Wu}. 

\begin{figure}
\centerline{\epsfxsize=8cm
            \epsfbox{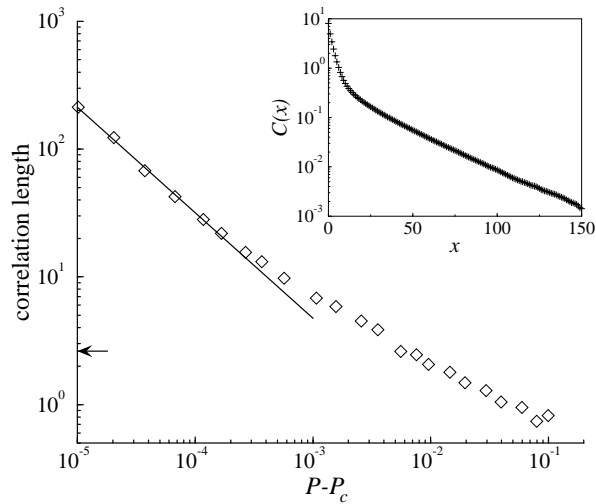}
            \vspace*{2mm}   }
\caption{Log-log plot of the correlation length characteristic to the
linear extension of alliances as a function of $P-P_c$. The solid line
represents the fitted power law behavior with a slope of -0.82.
The arrow points to the value of the correlation length characteristic
to the linear domain size inside an alliance for $P=0$.
The inset shows the lin-log plot of the correlation function
for $P=0.00055$.}
\label{fig:clexpfit}
\end{figure}

To obtain further evidence, we have also studied some other quantities
characterizing the critical behavior.
For example, the fluctuation of the order parameter defined as
$\chi = N \langle (m - \langle m \rangle)^2 \rangle$ can be well
approximated by a power law; $\chi \approx |P -P_c|^{\gamma}$
in the vicinity of $P_c$.
Below and above the critical point, numerical fitting yields 
$\gamma=1.3(2)$ and $\gamma^{\prime}=1.43(4)$ respectively, which values
agree with the theoretical prediction ($\gamma=\gamma^{\prime}=13/9$)
\cite{Wu}. The investigation of the cumulant of the order parameter
\cite{BH} for small lattice sizes ($L=60$, 100, 200) supports the presence
of a continuous transition at the critical mutation rate $P_c$.
Furthermore, we have determined the correlation function $C(x)$ for $P>P_c$,
which characterizes the probability of finding the same species on two sites
at a distance $x$ away from each other. In the vicinity of $P_c$, two different
characteristic lengths can be obtained from $C(x)$ (see the inset in 
Fig.~\ref{fig:clexpfit}). The shortest correlation length is proportional
to the linear size of a domain within the alliance, and this quantity
remains finite if $P \to P_c$. The longest correlation length is more
interesting, because it characterizes the linear size of the alliance
and diverges if $P \to P_c$. More precisely, this correlation length
can be well described by a power law ($\xi \sim (P-P_c)^{\nu}$)
as illustrated in Fig.~\ref{fig:clexpfit}. Numerical fitting predicts
$\nu = 0.82(4)$, in agreement with the theoretical prediction $\nu=5/6$
for the two-dimensional, three-state Potts model \cite{Wu}.

We have also investigated the model analytically, by evaluating
the probability of configurations on two adjacent sites (pair 
approximation method). Even though the number of possible pair
configurations is large ($9^2$), considering the (translation, rotation,
reflection, and cyclic) symmetries of the
system the number of different pair probabilities reduces to as few
as four. For sufficiently large $P$, this method gives a good
approximation for the behavior of the simulation model (i.e., it
predicts vanishing correlations if $P \to 1/2$). However, it shows
no sign of the phase transition found in the MC simulations at
$P_c$. This failure of the pair approximation method in showing the
phase transition is related to the key role that interfacial invasion
plays in the development of the self-organizing domain
structure.\cite{T94,SSM} This feature limits the techniques we can
use for further investigations.

In conclusion, our MC simulations justify that the present model exhibits
a critical phase transition accompanied with spontaneous symmetry-breaking,
in close analogy to the well-known Potts model. Here the mutation rate $P$
plays the role of the control parameter whose increase drives the system
towards the symmetric stationary state in which all the nine species are
present with the same probability. Conversely, for low mutation rate $P$ the
system drifts towards the dominance of an alliance consisting of three species
whose survival is maintained by cyclic within-domain invasion. Due to the very
symmetric topology of species dominance relations the system admits three
equivalent alliances. Numerical analysis of the critical behavior ($\beta$,
$\nu$, and $\gamma$ exponents) supports our conjecture that the observed
phase transition belongs to the universality class of the three-state Potts
model. The most surprising result of this work is that cyclic invasion is
capable of providing protection (stability) for alliances consisting of
mortal enemies (species double-dominating each other) under some particular
conditions hidden in the topology of the interaction. Further systematic
research is required to clarify the conditions for the emergence of such
defensive alliances accompanied by a reduction in the number of surviving
species for more general interaction topologies.

\acknowledgements
Support from the Hungarian National Research Fund (T-23552 and T-25793) is
acknowledged.


\begin{references}

\bibitem{bacik1}P. Reeves, {\it The Bacteriocins}
(Springer Verlag, New York, 1972).

\bibitem{bacik2}R. James, C. Lazdunski, and F. Pattus, (eds) {\it Bacteriocins, 
Microcins and Lantibiotics} (Springer Verlag, New York, 1991).

\bibitem{HS}J. Hofbauer and K. Sigmund, {\it Evolutionary Games and
Population Dynamics} (Cambridge University Press, Cambridge, 1998).

\bibitem{BG}M. Bramson and D. Griffeath, Ann.\ Prob.\ {\bf 17}, 26 (1989).

\bibitem{Lotka}A. J. Lotka, Proc.\ Natl.\ Acad.\ Sci.\ USA
{\bf 6}, 410 (1920).

\bibitem{Volterra}V. Volterra, {\it Lecon sur la Theorie Mathematique de la
Lutte pour la Vie} (Gouthier-Villars, Paris, 1931).

\bibitem{T94}K. Tainaka, Phys.\ Rev.\ E {\bf 50}, 3401 (1994).

\bibitem{FKB}L. Frachebourg, P. L. Krapivsky, and E. Ben-Naim, Phys.\ Rev.\
E {\bf 54}, 6186 (1996).

\bibitem{FK}L. Frachebourg and P. L. Krapivsky, J. Phys.\ A: Math.\ Gen.\
{\bf 31}, L287 (1998).

\bibitem{SSM}G. Szab\'o, M. A. Santos, and J. F. F. Mendes, Phys.\ Rev.\
E {\bf 60}, 3776 (1999).

\bibitem{Potts}R. B. Potts, Proc.\ Camb.\ Phil.\ Soc.\ {\bf 49}, 106 (1952).

\bibitem{Wu}F. Y. Wu, Rev.\ Mod.\ Phys.\ {\bf 54}, 235 (1982).

\bibitem{Binder}K. Binder, J.\ Stat.\ Phys.\ {\bf 24}, 69 (1981).

\bibitem{Grins}G. Grinstein, C. Jayaprakash, and Yu He, Phys.\ Rev.\
Lett.\ {\bf 55}, 2527 (1985).

\bibitem{BH}K. Binder and D. W. Heermann, {\it Monte Carlo Simulation
in Statistical Physics} (Springer-Verlag, Berlin, 1988).


\end{references}
\end{document}